\documentclass[pre,superscriptaddress,amsmath, amssymb,reprint]{revtex4-1}

\pdfoutput=1

\usepackage{graphicx}
\usepackage{epstopdf}
\usepackage[pdftex,breaklinks=true,bookmarksopen=true,bookmarksopenlevel=3,bookmarksnumbered=true,colorlinks=true,urlcolor= magenta,citecolor=blue,linkcolor=blue]{hyperref}

\usepackage{float} 

\newcommand{\cmc}{\,\mathrm{cm}^{-3}}

\newcommand{\mic}{\,\mu\mathrm{m}}

\begin{document}

\title[Betatron emission as a diagnostic]{Betatron emission as a diagnostic for injection and acceleration mechanisms in laser-plasma accelerators}

\author{S.~Corde}
\author{C.~Thaury}
\author{K.~Ta Phuoc}
\author{A.~Lifschitz}
\author{G.~Lambert}
\address{Laboratoire d'Optique Appliqu\'ee, ENSTA ParisTech - CNRS UMR7639 - \'Ecole Polytechnique, Chemin de la Huni\`ere, 91761 Palaiseau, France}
\author{O.~Lundh}
\address{Laboratoire d'Optique Appliqu\'ee, ENSTA ParisTech - CNRS UMR7639 - \'Ecole Polytechnique, Chemin de la Huni\`ere, 91761 Palaiseau, France}
\address{Department of Physics, Lund University, P.O. Box 118, S-22100 Lund, 8 Sweden}
\author{P.~Brijesh}
\address{Laboratoire d'Optique Appliqu\'ee, ENSTA ParisTech - CNRS UMR7639 - \'Ecole Polytechnique, Chemin de la Huni\`ere, 91761 Palaiseau, France}
\author{L.~Arantchuk}
\address{Laboratoire de Physique des Plasmas, \'Ecole Polytechnique-CNRS UMR7648, Route de Saclay, 91128 Palaiseau, France}
\author{S.~Sebban}
\author{A.~Rousse}
\author{J.~Faure}
\author{V. Malka}
\address{Laboratoire d'Optique Appliqu\'ee, ENSTA ParisTech - CNRS UMR7639 - \'Ecole Polytechnique, Chemin de la Huni\`ere, 91761 Palaiseau, France}

\begin{abstract}
Betatron x-ray emission in laser-plasma accelerators is a promising compact source that may be an alternative to conventional	x-ray sources, based on large scale machines. In addition to its potential as a source, precise measurements of betatron emission can reveal crucial information about relativistic laser-plasma interaction. We show that the emission length and the position of the x-ray emission can be obtained by placing an aperture mask close to the source, and by measuring the beam profile of the betatron x-ray radiation far from the aperture mask. The position of the x-ray emission gives information on  plasma wave breaking and hence on the laser non-linear propagation. Moreover, the measurement of the longitudinal extension helps one to determine whether the acceleration is limited by pump depletion or dephasing effects. In the case of multiple injections, it is used to retrieve unambiguously the position in the plasma of each injection. This technique is also used to study how, in a capillary discharge, the variations of the delay between the discharge and the laser pulse affect the interaction. The study reveals that, for a delay appropriate for laser guiding, the x-ray emission only occurs in the second half of the capillary: no electrons are injected and accelerated in the first half.
\end{abstract}

\maketitle

\section{Introduction}
The development of ultra short and bright x-ray sources  is one of the major axes of research in ultra fast phenomena science. These sources are of interest in many domains including material science, condensed matter, biology, chemistry and medicine. The development of compact x-ray sources using laser-plasma accelerators is a very active and promising field of research. Since the first observations of betatron radiation in the x-ray range in beam-driven plasma accelerators \cite{wang02} and in laser-plasma accelerators \cite{rous04} in the forced laser wakefield regime \cite{malk02}, impressive works have been done to characterize this source. 
The source size was measured using two different methods. The first method is based on Fresnel edge diffraction. The shadow of a knife edge that intercept the x-ray beam is observed onto a detector placed at a distance sufficient to ensure an appropriate resolution. The measurement in the detector plane of the intensity gradient near the edge of the shadow provides the spatial coherence and hence the source size~\cite{shah06}. Using this technique, it was demonstrated that the source size is less than 8 $\mic$, limited by the measurement resolution. The second method uses the strong correlations between the spatial distribution of the betatron radiation and the electron orbits. This method, discussed in details in references \cite{taph06, taph08}, led to an estimate of the source size of 2 $\mic$.
The spectrum of the betatron radiation was first measured using a set of filters \cite{rous04,taph05} and then using an imaging spectrometer consisting of a toroidal mirror and a Bragg crystal \cite{albe08}. The spectrum measured  at Laboratoire d'Optique Appliqu\'ee (LOA) has a typical wiggler shape with a critical energy of about 1 keV, a divergence of about $10$ mrad with $10^{5}$ photons/0.1\% bandwidth/shot  at 1 keV and a total number of photons in the range $10^{8}-10^{9}$.
The betatron pulse duration was estimated using time-resolved x-ray diffraction. Ultra-fast phase transition (non-thermal melting of InSb) was used as a Bragg switch to sample the x-ray pulse duration \cite{taph07}. The duration of the betatron x-ray pulse was estimated to be less than 1 ps with a best fit below 100 fs. A more recent experiment indicates that the duration may be of a couple of femtoseconds~\cite{lund11}. Using a PW class, 300 J - 600 fs laser system at Rutherford Appleton Laboratory, in a regime for which the laser pulse is much longer than the plasma wavelength, betatron radiation with higher critical energy of about $30$ keV and a divergence of about $1$ rad was measured \cite{knei08}. Betatron radiation was also  observed in an experiment performed on the Michigan Hercules laser system at the University of Michigan \cite{knei09}. The critical energy was about 10 keV and the divergence angle of about 10 mrad. The source size was measured to be $\approx 1 \mic$~\cite{knei10}. More recently, betatron radiation was generated with controlled features using electrons produced by colliding pulse injection \cite{faur06}, showing clear correlations between the electron beam and the betatron radiation \cite{cord11a}. 

We report in this article on the measurement of the longitudinal position and emission length of the x-ray emission. The method relies on the observation of the shadow of an aperture mask, adequately situated close to the source, in the betatron x-ray beam profile. The size of the object shadow on the x-ray image allows to determine the x-ray emission longitudinal position in the plasma, while the intensity gradient of the edge of the shadow yields the emission length. The intensity gradient of the edge is dominated by the longitudinal extension of the source, and not by the transverse extension, in contrast to other studies on the x-ray source size \cite{shah06, knei10}. Because the x-ray emission position and length are closely connected to the electron injection position and the acceleration length, this measurement provides an insight into the interaction. The paper is structured as followed. In section 2, we describe briefly the principle of betatron emission in laser-plasma accelerators. Section 3 introduces the method that is used to measure the longitudinal position and extension of the x-ray emission in laser-plasma accelerator. In section 4, we show how the betatron emission profile can be used to infer properties of electron injection and acceleration in gas cell experiments. Then, in section 5, we present a study of the laser-plasma interaction in a capillary discharge accelerator that show the potential and the usefulness of this diagnostic for such type of experiment. Conclusions are presented in section 6.

\section{Principle of betatron emission in laser-plasma accelerators}

\begin{figure}
\begin{center}
\includegraphics[width=8.5cm]{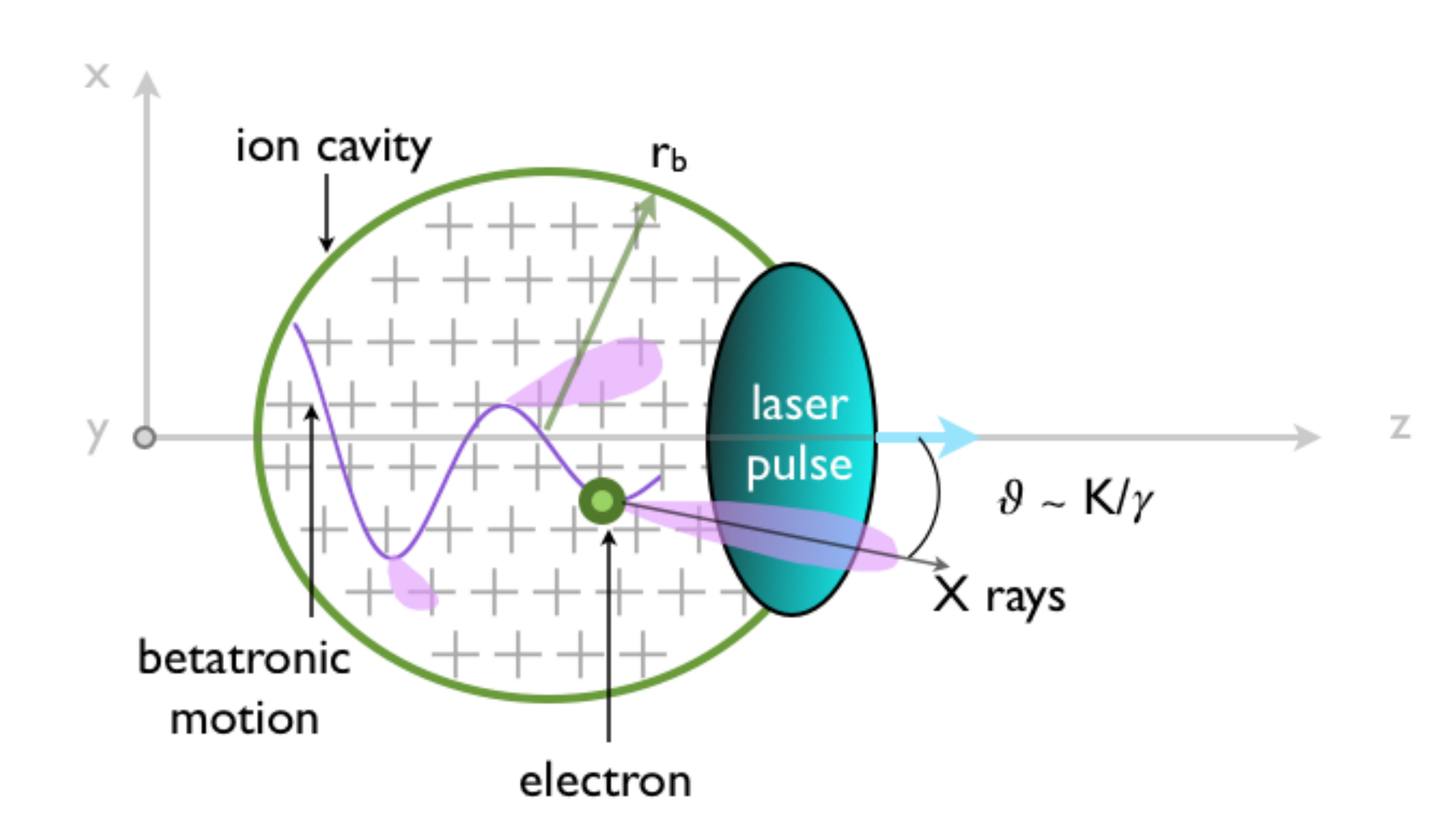}
\caption{Principle of betatron emission in a laser-plasma accelerator.}
\label{BetatronPrinciple}
\end{center}
\end{figure}

An intense laser pulse propagating in an under-dense plasma can drive a relativistic plasma wave in which electrons can be injected and accelerated to relativistic energies. To date, the most efficient mechanism to accelerate electrons in a plasma wave is called the bubble, blow-out or cavitated wakefield regime~\cite{pukh02, lu06, lu07}. In that regime, the wake consists of a spherical ion cavity as represented in Fig.~\ref{BetatronPrinciple}. This regime is reached if the waist $w_0$ of the focused laser pulse is close to the plasma wavelength  and if the pulse duration is approximately equal to half a plasma wavelength. In addition, the laser intensity must be sufficiently high $(a_0 > 2)$ to expel most of the electrons out of the focal spot. When these conditions are met, an ion cavity is formed in the wake of the laser pulse. Electrons can be trapped at the back of the cavity and accelerated by the high electric field with an accelerating force of the order of 100 GeV/m.
While the electron motion is essentially longitudinal, along the laser propagation axis, electrons experience as well a transverse force.  An electron, injected  off-axis, is accelerated in the longitudinal direction $x$ and oscillates across the cavity axis at a period called the betatron period. 

In this interaction regime, synchrotron-like radiation, called betatron radiation, is naturally produced~\cite{esar02, kost03}. It is a moving charged particle radiation emitted by relativistic electrons oscillating in the ion cavity. The features of the betatron radiation directly depend on the electron orbits. The radiation is emitted in the direction of the electron velocity. The spectrum depends on the period and on the transverse amplitude of the electron orbits. As an example, the betatron radiation produced at the interaction of $10-100$~TW class laser with a millimeter scale helium gas jet at density of the order of $10^{19}$ cm$^{-3}$ consists of a beam collimated within a few tens mrad, containing about $10^9$ photons integrated over a broadband spectrum extending up to about 10 keV~\cite{four11a, four11b}. The pulse duration is femtosecond~\cite{lund11}, the source size is of the order of a micrometer or less~\cite{taph06, taph08, albe08, knei10, cord11a, four11b, mang09, schn12, plat12} and the peak brightness is of the order of $10^{22}$ ph/0.1$\%$ BW/s/mm$^2$/mrad$^2$~\cite{knei10}. 
 
\section{Measuring the longitudinal position and extension of the x-ray emission} 

Betatron radiation can be used to determine some features of the interaction and of the acceleration process in a laser-plasma accelerator. For example, the radiation being emitted in the direction of the electron velocity, the x-ray beam profile is an image of the transverse orbits of the electrons in the wakefield~\cite{taph06}. Here, we will use the betatron radiation to obtain information on the laser-plasma interaction in the wakefield cavity.

\begin{figure}
\begin{center}
\includegraphics[width=8.5cm]{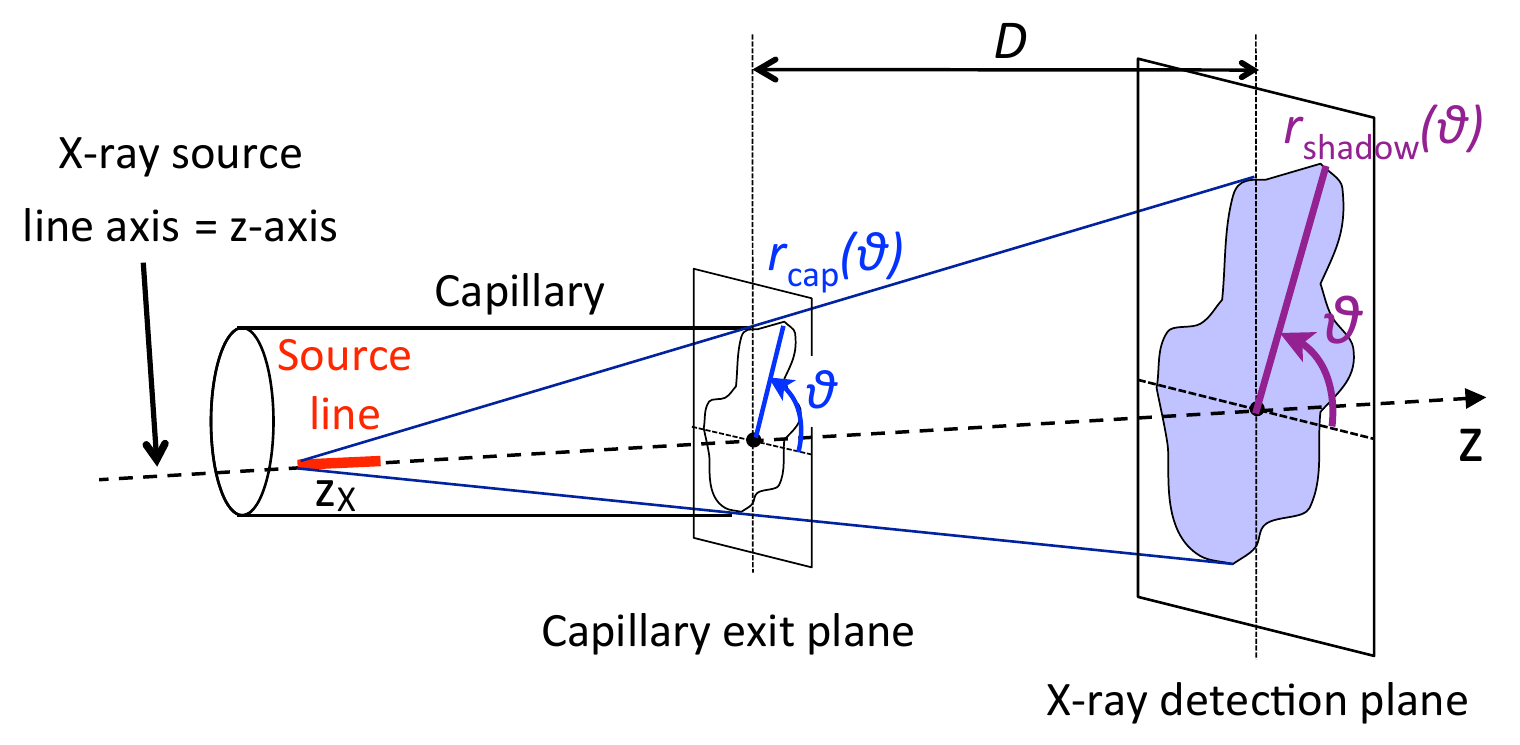}
\caption{Schematic illustrating the principle of the method and how the $z$-axis, and the functions $r_\mathrm{cap}(\theta)$ and $r_\mathrm{shadow}(\theta)$ are defined.}
\label{princ}
\end{center}
\end{figure}

The method relies on the measurement of the position and the longitudinal extension of the betatron x-ray emission. To do this, the x-ray shadow of an aperture mask positioned close to the source is measured. The principle of the method is shown on Fig.~\ref{princ}. The size of the aperture shadow on the x-ray image allows one to determine the x-ray emission longitudinal position in the plasma, while the intensity gradient of the edge of the shadow yields the emission length. Because the x-ray emission position and length are closely connected to the electron injection position and the acceleration length, this measurement provides an insight into the interaction.

The experiment was conducted at Laboratoire d'Optique Appliqu\'ee with the ``Salle Jaune'' Ti:Sa laser system, which delivers 0.9 J of laser energy on target with a full-width at half-maximum (FWHM) duration of 35 fs and a linear polarization. The laser pulse was focused by a 1 m focal length spherical mirror at the entrance of a capillary. The FWHM focal spot size was 22 $\mu$m, and using the exact intensity distribution in the focal plane we found a peak intensity of $3.2\times10^{18}\:\textrm{W.cm}^{-2}$, corresponding to a normalized amplitude of $a_0=1.2$.
The target was a capillary with a diameter of $d_\mathrm{cap} = 210\,\mu$m and a length of 15~mm, filled with hydrogen gas whose backing pressure ranges from 50 to 500 mbar.  The x-ray beam profile was measured using a  x-ray CCD camera with $2048\times2048$ pixels of size $13.5\:\mu\textrm{m}\times13.5\:\mu\textrm{m}$, situated at $D=73.2$~cm from the capillary exit and protected from the laser light by a 20 $\mu$m Al filter. Electrons were characterized using a focusing-imaging spectrometer~\cite{rech09a}.

\begin{figure*}
\begin{center}
\includegraphics[width=0.8\textwidth]{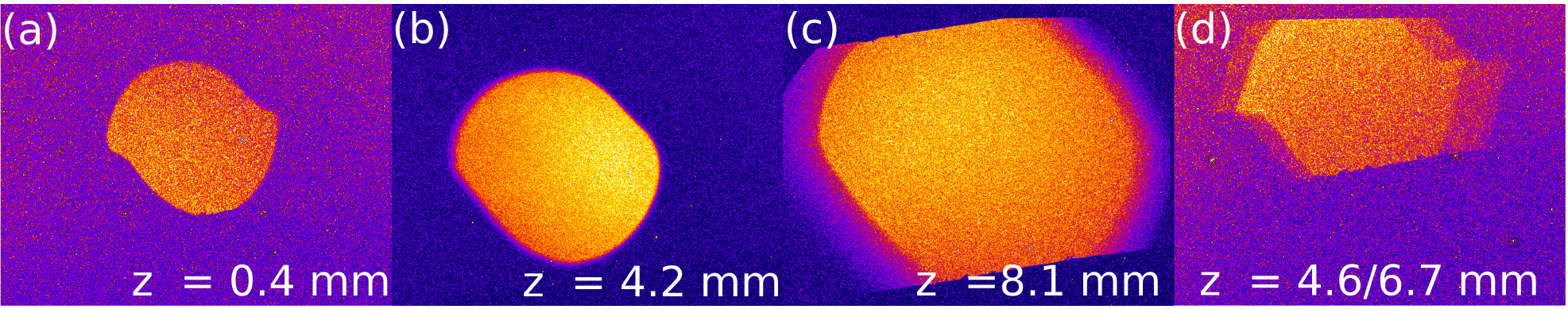}
\caption{(a)-(d) X-ray beam profiles measured for different emission positions in the plasma. The imprint of the capillary exit in the x-ray beam profile allows one to deduce the x-ray emission longitudinal position $z_X$ in the plasma. Each x-ray image has a $2.76\:\textrm{cm}\times2.13\:\textrm{cm}$ size, and the camera is situated at $D=73.2$ cm from the capillary exit. The x-ray image (d) shows two different emission positions in the plasma, respectively at 4.6 and 6.7~mm from the entrance.}
\label{resu}
\end{center}
\end{figure*}

Because the betatron emission has a divergence larger than the opening angle associated to the capillary exit,  the exit acts as an aperture mask that clips the x-ray beam~\cite{geno11}. This is illustrated in Fig.~\ref{resu}, which displays different x-ray beam profiles measured during the experiment. The shape of the capillary exit (the capillary is made of two sapphire plates with half-cylindrical grooves, which are slightly misaligned here), is visible in the x-ray images, but with different sizes. Depending on the longitudinal position of the x-ray source, $z_X$, the capillary exit shadow size varies on the camera, because the opening angle associated to the capillary exit changes. Hence, the measured x-ray profiles allow us to determine the longitudinal position of the x-ray source in the capillary.
If x-rays were emitted from a point source, the edge of the shadow will be perfectly sharp, while for a finite source size, the edge  presents a finite gradient, which depends on the transverse and longitudinal extension of the x-ray source. In previous experiments, the transverse source size of the x-ray source was measured to be on the order of  $1 - 2 \mic$ or less~\cite{taph06, taph08, albe08, knei10, cord11a, four11b, mang09, schn12, plat12}. For our experimental set-up, we found that all x-ray images present gradients much larger than those induced by a transverse size of $1-2$ microns (for $z_X=5$~mm, a transverse source size of $1 \mic$ gives the same gradient as a longitudinal extension of $100 \mic$), and therefore the gradient length is dominated by the longitudinal extension of the source. X-rays can thus be considered as being emitted by a longitudinal source line, and the measurement of the gradient length in the x-ray images yields the extension of this x-ray source line.
In the following, we use a cylindrical coordinate system $(r,\theta, z)$ whose $z$-axis is the source line axis. If $z_{\mathrm{entrance}}=0$ corresponds to the entrance of the capillary and $z_{\mathrm{exit}}=15$~mm to the exit, then the x-ray emission position is given for $r_\mathrm{cap}(\theta) \ll r_\mathrm{shadow}(\theta)$ by $z_X \simeq z_\mathrm{exit}-r_\mathrm{cap}(\theta)D/r_\mathrm{shadow}(\theta)$, where $r_\mathrm{cap}(\theta)$ [respectively $r_\mathrm{shadow}(\theta)$] is the radial distance between the $z$-axis and the capillary edge (respectively the shadow edge) in the direction defined by the angle $\theta$ (see Fig. \ref{princ}), and $D$ is the distance between the capillary exit and the observation plane. For a perfectly circular capillary exit and a line source on the capillary axis, $r_\mathrm{cap}(\theta)$ simplifies to $d_\mathrm{cap}/2$, but a more general capillary exit shape and an arbitrary position or orientation of the line source can be represented by the function $r_\mathrm{cap}(\theta)$. 

 Assuming the betatron x-ray beam profile without the mask is constant on the gradient scale length (a reasonable approximation for our experimental results), the signal profile reads:
\begin{equation}
S(r, \theta)=\int_{z(r, \theta)}^{z_{\mathrm{exit}}}\frac{dI(z^\prime)}{dz^\prime}dz^\prime,
\label{eq1}
\end{equation}
for $z(r, \theta) = z_{\mathrm{exit}}-r_{\mathrm{cap}}(\theta)D/r\in\left[z_{\mathrm{entrance}},z_{\mathrm{exit}}\right]$. In Eq.~(\ref{eq1}), $dI(z^\prime)$ is the x-ray signal originated from the emission between $z^\prime$ and $z^\prime+dz^\prime$, $S(r, \theta)$ is the signal measured at a given position $(r, \theta)$ on the detector and $r_{\mathrm{cap}}(\theta)$ is the radial distance between the $z$-axis and the capillary edge in the direction defined by the angle $\theta$ (see Fig.~\ref{princ}).
Equation (\ref{eq1}) can be understood as follow. For a position $(r_0, \theta_0)$ on the detector, rays coming from $z^\prime<z(r_0, \theta_0)$ are blocked by the capillary exit, and therefore the signal measured at $(r_0, \theta_0)$ is the sum of the signal emitted between $z(r_0, \theta_0)$ and $z_{\mathrm{exit}}$. Taking the derivative of Eq. (\ref{eq1}), the longitudinal profile of the x-ray emission $dI(z)/dz$ can be expressed as a function of the signal radial profile in the detector plane $S(r, \theta)$:
\begin{equation}
\frac{dI(z)}{dz}=-\frac{\partial S(r(z, \theta), \theta)}{\partial r}\frac{r(z, \theta)^2}{r_{\mathrm{cap}}(\theta)D} ,
\label{eq2}
\end{equation}
where $r(z, \theta)=r_{\mathrm{cap}}(\theta)D/(z_{\mathrm{exit}}-z)$. If $\delta z$ is the characteristic emission length and $\delta r(\theta)$ the characteristic intensity gradient length, then $\delta z=\delta r(\theta)(z_{\mathrm{exit}}-z_X)^2/(r_{\mathrm{cap}}(\theta)D)$ for $\delta z/(z_{\mathrm{exit}}-z_X)\ll1$. This implies that $r_{\mathrm{cap}}(\theta)\propto \delta r(\theta)$. As a consequence, the measurement of the intensity gradient in the image plane $\delta r(\theta)$ yields the longitudinal length of the x-ray emission, $\delta z$. The full emission profile $dI(z)/dz$ can be retrieved from $\partial S/\partial r$ using Eq.~(\ref{eq2}). Lastly, the transverse displacement of the shadow and the asymmetry of $\delta r(\theta)$ provide information on the orientation and transverse position of the source line. For example, we observed during an experimental run a vertical low drift of the line source which was correlated with a low vertical drift of the laser pulse. The asymmetry observed on some shots in $\delta r(\theta)$ [see Fig.~\ref{resu}(b)] also confirms that the large intensity gradients observed in Fig.~\ref{resu} are not originated from the transverse source size of the betatron emission, since a transverse extension can only lead to symmetrical intensity gradients.

\section{Application to a gas cell laser-plasma accelerator}

\begin{figure*}
 \begin{center}
\includegraphics[width=0.8\textwidth]{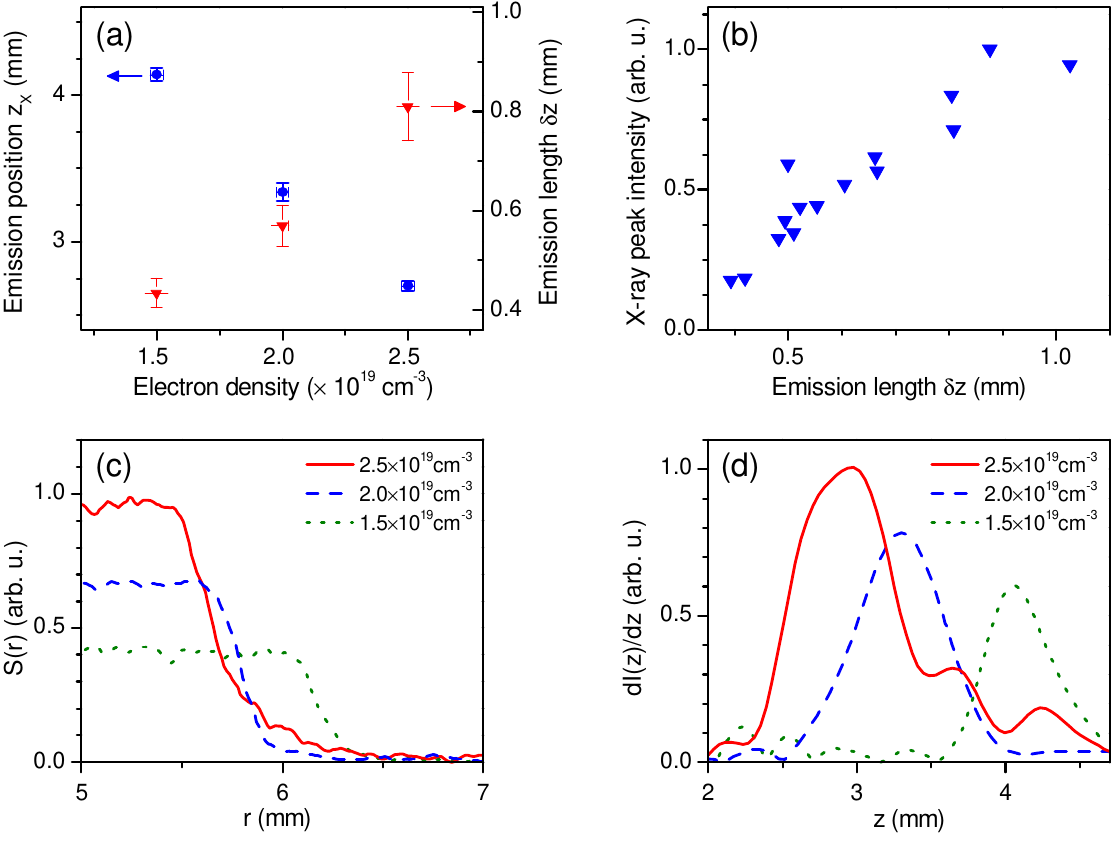}
 \caption{Experimental results obtained in a steady-state-flow gas cell laser-plasma accelerator. (a) Position of the beginning of the x-ray emission region $z_{X}$ (blue squares), and emission length $\delta z$ (defined such that 70\% of the signal is emitted in $\delta z$) (red triangles), as a function of the electron density. Each measurement point corresponds to an average over 5 to 7 shots, and the vertical error bars give the standard error of the mean (statistical error). For the emission length $\delta z$, neglecting the transverse extent of the source gives a systematic error that is estimated to be approximately equal or smaller than the represented statistical error. (b) X-ray peak intensity as a function of $\delta z$. (c) Examples of raw lineouts of the edge intensity profiles $S(r)$. (d) Corresponding single shot x-ray emission longitudinal profiles $dI(z)/dz$.}
 \label{figu}
 \end{center}
\end{figure*}

We studied the influence of the plasma electron density $n_e$ on the x-ray emission position $z_X$ and longitudinal extension $\delta z$, using the capillary as a steady-state-flow gas cell. No x-rays are observed for electron density below $1.5\times 10^{19} \cmc$, and the x-ray signal is increasing from the threshold at $1.5\times 10^{19} \cmc$ up to $2.5\times 10^{19} \cmc$. In these conditions, we observed broadband electron beams with energies from 100 to 400 MeV, with sometimes some mono-energetic components, and charge in the few tens of pC range. Figure \ref{figu}(a) shows the behavior of $z_X$ and $\delta z$ with respect to $n_e$. 
The position $z_X$ of the beginning of the x-ray emission varies from 4.1~mm to 2.7~mm when $n_e$ increases from $1.5\times 10^{19} \cmc$ to $2.5\times 10^{19} \cmc$. This behavior can be understood by the modification of the laser propagation in the plasma. When the density increases, the laser pulse self-focuses and self-steepens more quickly and towards a smaller transverse spot size~\cite{esar97}.
As a result, it attains sufficiently large $a_0$ to trigger electron trapping in a smaller propagation distance. Moreover, electron self-injection is facilitated at high density, due to the stronger wakefield amplitude and the reduced wake velocity and wave-breaking threshold, which could also contribute to an x-ray emission beginning sooner for high density.

The x-ray peak intensity is plotted  in Fig.~\ref{figu}(b)  as a function of the emission length $\delta z$ (in this figure the electron density is not constant). The x-ray signal is clearly increasing with the emission length, in a nearly linear way. It shows that one of the key parameters for increasing the x-ray signal in our configuration is actually the emission length.  Figure~\ref{figu}(d) shows that the peak of  $dI(z)/dz$ depends weakly on $n_e$. As a result, variations of the emission length are the main source of x-ray signal changes. This conclusion is supported by the fact that in the experiment, the electron peak energy was observed to be a weak function of $n_e$.

Figure~\ref{figu} also shows that the emission length $\delta z$ depends on the electron density $n_e$. It increases from 430 $\mu$m to 810 $\mu$m when the electron density varies from $1.5\times 10^{19} \cmc$ to $2.5\times 10^{19} \cmc$. Further, at high density, the x-ray emission length extends well beyond the dephasing and depletion lengths (the orders of magnitude are respectively $L_d\sim 200\mic$ and $L_\mathrm{pd}\sim500\mic$ for $n_e=2.5\times10^{19}\cmc$ considering Lu's model~\cite{lu07}). This is counter-intuitive since we expect the interaction to finish faster and the total interaction distance and the emission length to be reduced at higher density, because both the dephasing and depletion lengths decrease with $n_e$. A possible explanation for that experimental observation relies on the transition from a laser wakefield accelerator (LWFA) to a plasma wakefield accelerator (PWFA) \cite{pae10} in which the wakefield is excited by a particle beam \cite{blum07}. At higher density, the laser pulse amplitude $a_0$ attains a higher value, allowing higher amplitude plasma wakefields and stronger electron self-injection. A sufficiently dense injected and accelerated electron beam can pursue the wakefield excitation after the laser pulse has depleted,  increasing the total interaction distance and the x-ray emission length. Since this mechanism needs dense electron beams, it will be facilitated at higher density where the injection is stronger.

\begin{figure*}
 \begin{center}
\includegraphics[width=\textwidth]{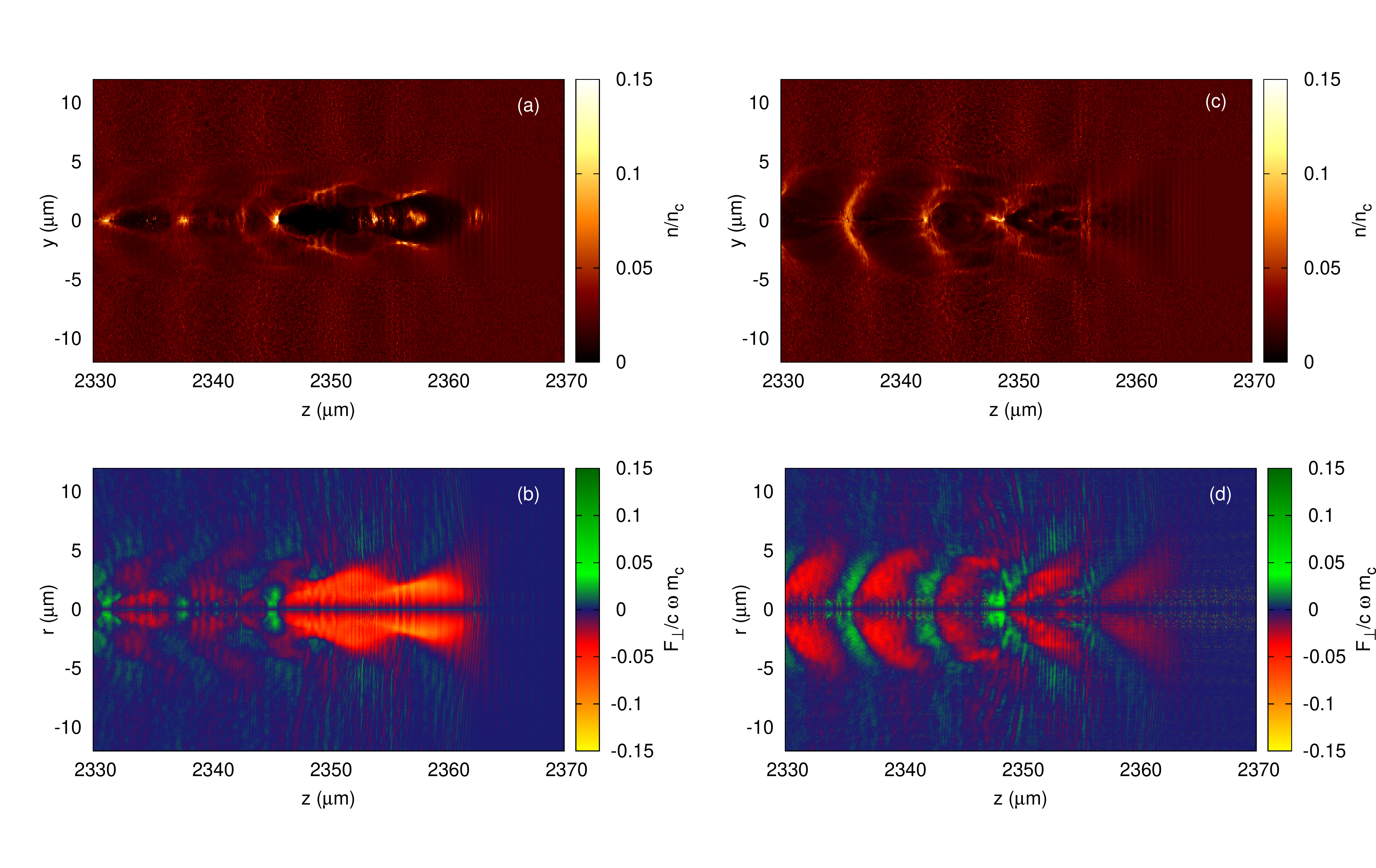}
 \caption{Simulated laser wakefield at the end of the laser-plasma interaction for $n_{e} = 2.5\times 10^{19} \cmc$. (a)-(c) Electron density $n_e$, normalized to the critical density $n_c =
m_e\epsilon_0\omega^2/e^2$, where $\omega$ is the laser frequency. (b)-(d) Transverse force $F_{\perp}= -e(E_{r}-cB_{\theta})$, normalized to $m_ec\omega$. In (c)-(d), the influence of the electron beam is removed (the laser pulse is extracted and re-injected in a homogeneous plasma in order to calculate
the wakefield induced by the laser pulse only).}
 \label{simu}
 \end{center}
\end{figure*}

To check this scenario, we performed Particle-In-Cell simulations with the  Calder-Circ code~\cite{lifs09}. This code uses a Fourier decomposition of the electromagnetic fields in the azimuthal direction. The first two modes are retained, which allows us to describe the linearly polarized laser field and a quasicylindrical wakefield. The normalized laser amplitude was $a_0 = 1.1$, the FWHM focal spot width was $22 \mic$ and the FWHM pulse duration was 35 fs. We simulated the high density case, $n_e=2.5\times10^{19}\cmc$, and found a similar x-ray emission longitudinal profile as in the experiment~\cite{cord11b}, with an emission extending well beyond the depletion length. At a late time, where the laser amplitude is strongly reduced (from a maximum of $a_0\gtrsim4$ to $a_0\lesssim2$), x-ray emission still occurs. Figures \ref{simu}(a) and \ref{simu}(b) show the simulated wakefield at this late time, where the ionic cavity (a) and the strong focusing force (b) applied to the electrons are visible. To verify that, at this late time, the wakefield is mainly excited by the electron beam itself and not anymore by the laser pulse, we simulated the wakefield excited by the laser pulse only. To do so, the laser pulse was extracted from the simulation and re-injected in a homogeneous plasma. Figures \ref{simu}(c) and \ref{simu}(d) show the result: the laser pulse is unable to excite a strong transverse wakefield, and the wakefield obtained is negligible compared to the one excited in the presence of the electron beam [compare Figs.~\ref{simu}(b) and~\ref{simu}(d)]. We can therefore conclude that, at this late time where the laser amplitude is strongly reduced, the wakefield is mainly excited by the electron beam. As a result, a transverse wakefield is maintained by the electron beam such that electrons continue to oscillate and to emit x-rays. This explains why x-ray emission is not limited by the dephasing length nor by laser depletion.
The increase in $\delta z$ with $n_{e}$, observed in Fig.~\ref{figu}(a), could be explained by a higher normalized particle beam density $n_{p}/n_{e}$ at higher density, favoring an electron beam excited transverse wakefield and a late x-ray emission.

During the experiment, multiple emission positions were observed on some shots, as shown for instance in Fig. \ref{resu}(d), where x-rays are emitted at $z_{X}=4.6$ and $z_{X}=6.7$~mm. This can be explained by oscillations of the laser pulse amplitude $a_{0}$ during its propagation in the plasma. The wakefield amplitude is sufficiently high to trap electrons only when $a_{0}$ is at its maximum, leading to multiple electron injection and therefore multiple emission positions.

\section{Application to a capillary discharge laser-plasma accelerator} 

\begin{figure}
  \begin{center}
\includegraphics[width=8.5cm]{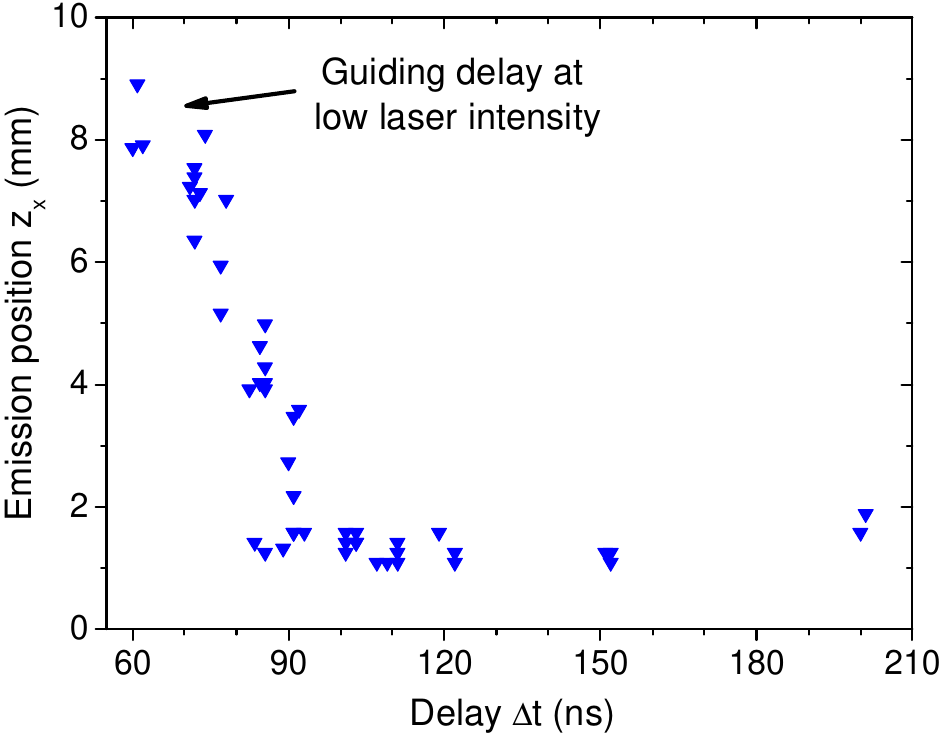}
\caption{Position $z_X$ of the beginning of the x-ray emission as a function of the delay between the electrical discharge and the laser pulse, in a capillary discharge laser-plasma accelerator. The hydrogen gas backing pressure was set to 300 mbar.}
    \label{dech}
    \end{center}
\end{figure}

The method for measuring $z_X$ and $\delta z$ was also applied to the case of a capillary discharge laser-plasma accelerator~\cite{leem06, naka07, kars07, rowl08, ibbo10a, ibbo10b}. In that case, we triggered an electrical discharge by applying a high voltage at the edges of the capillary~\cite{spen00, butl02}. The charging voltage of the 2~nC capacitor was set to 25~kV (see Ref.~\cite{rech09b} for details regarding the electrical discharge circuit design), and the resulting current pulse had a peak of $\sim600$~A and a duration of $\sim200$~ns (full width). Heated on the capillary axis, the plasma cools down by the wall. After plasma expansion, this inhomogeneous temperature results in transverse density and index profiles which can guide the laser pulse over the entire capillary length. The study of the laser pulse guiding at a low laser intensity as a function of the delay between the discharge and the laser pulse revealed that the channel index profile evolves significantly with this delay. We found that laser guiding was possible for discharge triggered  $\Delta t=70\pm10$ ns before the laser enters the capillary. More precisely, we measured that, for this delay, 90\% of the laser energy is transmitted by the capillary and the focal spot size at the capillary exit varies by less than 10\% compared with the focal spot at the entrance. At full laser energy, electrons and x-rays could be produced either using a guiding delay or using discharge triggered sooner by a few tens of ns ($\Delta t > 90$~ns). The measurement of the x-ray emission position $z_X$ as a function of the delay $\Delta t$ between the electrical discharge and the laser pulse arrival is represented in Fig. \ref{dech}, for a hydrogen gas backing pressure of 300 mbar.

The x-ray emission position is totally different depending on the delay. When the discharge was triggered a few tens of ns sooner ($\Delta t > 90$~ns) than the guiding delay, x-rays were emitted on the very beginning of the capillary, indicating that only a short distance was used for laser-plasma interaction and electron/x-ray generation. On the other hand, when the delay becomes close to the guiding delay, the emission position drifts further inside the capillary, reaching $8-9$~mm with respect to the capillary entrance, which indicate that the laser pulse is guided, in agreement with the measurement at low laser energy. At lower backing pressure (150 and 225 mbar) and at the guiding delay, we observed the entire betatron beam profile without any capillary exit shadow, indicating an x-ray emission position $z_X$ situated between 9 and 15~mm from the capillary entrance. In that case, no transition between guided and non-guided configuration was observed, either the emission position $z_X$ was at $1-2$~mm (for soon discharge trigger), either it was between 9 and 15~mm (guiding delay).

Because of the small amount of betatron x-rays produced in this experiment, the x-ray images were too noisy to perform a precise analysis of the emission length as a function of the delay. We observed, however, that under any conditions (pressure, delay etc.), the emission length $\delta z$ was smaller than 1~mm. In other words, the acceleration length does not increase significantly even though the laser pulse is guided in the capillary. This is consistent with the experimental observation of a peak electron energy that does not depends strongly on the backing pressure nor on the delay. 

A possible explanation is that at high laser intensity a large part of the laser energy is lost during the  coupling of the laser into the waveguide. In this scenario, high laser intensity and non-linear effects are assumed to be responsible for the bad coupling. Indeed, our measurements at very low laser intensity showed a very good coupling [with more than 90\% of the energy transmitted through the capillary, and the laser pulse at the exit having the same transverse size as at the entrance (within 10\% precision)]. The index profiles at the capillary ends could be non-trivial and, while not affecting the laser pulse at low intensity, they could induce a loss of energy when at full laser energy with non-linear effects enabled. Because of this bad coupling at high intensity, a longer propagation is required for the laser to reach an intensity sufficient to drive a non-linear plasma wave and trigger the electron injection, which may explain why the x-ray emission occurs only in the second part of the capillary. Further, the laser energy available for the acceleration is reduced, which hinders the increase of the acceleration length (the laser cannot drive a plasma wave over a long distance).

Another explanation could be that the experiment operated at a too high plasma density (the backing pressure was always above 150 mbar). Indeed, the acceleration length may be increased by reducing the backing pressure (the dephasing and depletion lengths increase when $n_e$ is lowered). However, the electrical discharge in this experiment was unstable for backing pressure below 150 mbar, preventing any guiding at low plasma density.

\section{Conclusion}

Betatron radiation has been shown to be a very promising source for applications and a very interesting non-invasive tool to diagnose laser-plasma interaction. The method presented in this paper  allows the measurement of single shot x-ray emission longitudinal profiles in a laser-plasma accelerator. The method provides detailed information on the interaction that is of crucial importance to understand many experiments performed when an intense laser pulse is focused in a gas cell or a capillary discharge. 

In the case of a gas cell laser-plasma accelerator, we showed that, at a high density, x-ray emission begins sooner because of the faster self-focusing and self-steepening of the laser pulse. Also, our measurements showed a very large emission length at high density, which can be explained by including the role of the electron beam driven wakefield in the laser-plasma interaction picture. 

One of the major goals for laser-plasma accelerators consists in increasing the acceleration length, either by guiding the laser pulse or using higher laser energy. In this context, this mapping method should be a powerful tool for optimizing the acceleration. In this paper, we showed that the use of this method in the study of a capillary discharge laser-plasma accelerator can reveal very interesting behaviors. Electron acceleration and x-ray emission can occur for non-guiding delays, but the interaction takes place only in the first few millimeters of the 15 mm long capillary. In contrast, at the guiding delay, electron acceleration and x-ray emission begins in the second half and extends over a millimeter or less, showing some guiding effects but, surprisingly, no injection and acceleration occurs in the first half of the capillary.

This method can also be applied with gas jets by using a small aperture near the source, and will allow one to understand over which distance self-focusing and self-steepening take place, where electron injection occurs, and over which distance acceleration and x-ray emission happen. 

\section*{Acknowledgements}
The authors thank the Agence Nationale pour la Recherche, through the COKER project ANR-06-BLAN-0123-01, the European Research Council through the PARIS ERC project (under Contract No. 226424) and the support from EC FP7 LASERLABEUROPE/ LAPTECH Contract No. 228334 for their financial support. The authors also appreciate the contributions of J. Larour, P. Auvray, and S. Hooker in the realization of the capillary unit.

\end{document}